\documentclass[aps,prm,showpacs,superscriptaddress,a4paper,twocolumn]{revtex4-2}
\usepackage{amsmath}
\usepackage{graphicx} 
\usepackage{booktabs}
\begin{document}	
\title{Efficient Electrochemical CO$_2$ Reduction Reaction over Cu-decorated Biphenylene}	

\author{Radha N Somaiya}
\affiliation{Materials Modeling Laboratory, Department of Physics, IIT Bombay, Powai, Mumbai 400076, India}

\author{Muhammad Sajjad}
\affiliation{Nottingham Ningbo China Beacons of Excellence Research and Innovation Institute, University of Nottingham, Ningbo, China}
\affiliation{Key Laboratory of Carbonaceous Wastes Processing and Process Intensification Research of Zhejiang Province, University of Nottingham Ningbo China, Ningbo 315100, China}
		
\author{Nirpendra Singh}
\email{Nirpendra.Singh@ku.ac.ae}
\affiliation{Department of Physics, Khalifa University, Abu Dhabi, 127788, United Arab Emirates}
\affiliation{Research and Innovation Center for Graphene and 2D materials (RIC2D), Khalifa University, Abu Dhabi, United Arab Emirates}

\author{Aftab Alam}
\email{aftab@iitb.ac.in}
\affiliation{Materials Modeling Laboratory, Department of Physics, IIT Bombay, Powai, Mumbai 400076, India}

\begin{abstract}	
Developing efficient electrocatalysts for CO$_2$ reduction into value-added products is crucial for the green economy. Inspired by the recent synthesis of Biphenylene (BPH), we have systematically investigated pristine, defective, and Cu-decorated BPH as an electrocatalyst for the CO$_2$ reduction reactions (CRR). Our first-principles calculations show the CO$_2$ molecules weakly interact with the pristine BPH surface while defective BPH facilitates the CO$_2$ adsorption with a binding energy ($E_b$) of -3.22 eV, indicating the detrimental process for the CRR on the surface of both systems. Furthermore, we have investigated the binding energy and kinetic stability of Cu-decorated BPH as a single-atom-catalyst (SAC). The molecular dynamics simulations confirm the kinetic stability, revealing that the Cu-atom avoids agglomeration under low metal dispersal conditions. The CO$_2$ molecule gets adsorbed horizontally on the Cu-BPH surface with $E_b$ of -0.52 eV. The CRR mechanism is investigated using two pathways beginning with two different initial intermediate states, formate ($\mathrm{^*OCOH}$) and the carboxylic ($\mathrm{^*COOH}$) pathways. The formate pathway confirms the conversion of $\mathrm{^*OCOH}$ to $\mathrm{^*HCOOH}$ with the rate-limiting potential ($U_L$) of 0.57 eV for the production of HCOOH, while for the carboxylic pathway, the conversion of $\mathrm{^*COH}$ to $\mathrm{^*CHOH}$ has $U_L$ of 0.49 eV for the production of CH$_3$OH. We have also investigated the effect of protons using charged hydrogen pseudopotential, which hints towards the possible formation of CH$_3$OH as fuel. Our findings propose Cu-BPH as an efficient single-atom catalyst for CO$_2$ conversion compared to the well-known Cu metal.		
\end{abstract}

\maketitle	
\clearpage
\section{Introduction}
Energy consumption is increasing enormously, relying mainly on burning fossil fuels, which raises greenhouse gases in the environment. Due to the greenhouse effect, the CO$_2$ concentration has risen from 280 ppm in the early 1800s to 421 ppm till now \cite{li2011carbon,lu2022recent,orr2009co}. Photo(electro)chemical carbon dioxide reduction reactions (CRR) could be a promising route for reducing the greenhouse effect by converting CO$_2$ into value-added hydrocarbons\cite{maginn2010co2,whipple2010prospects}, which is challenging because CO$_2$ is a highly inert and linear molecule that requires dissociation energy of 750 kJ/mole \cite{zhang2017nanostructured}. In addition, the high overpotential is another major obstacle to poor selectivity. Therefore, efficient catalysts are required to steer the reaction towards value-added products having a low overpotential \cite{zhu2019carbon,kortlever2015catalysts}. These products include the formation of multiple hydrocarbons at various electron-proton transfer steps like carbon monoxide (CO), formic acid (HCOOH), methanol ($\mathrm{CH_3OH}$), and methane ($\mathrm{CH_4}$) via a two-, six-, and eight-electrons reduction pathway, respectively. The product selectivity strongly depends on intrinsic electronic and extrinsic physical properties, which determine the activation energy at each reaction step and the adsorption energies of intermediates \cite{jiao2017molecular}. This complexity makes designing value-added products even more complex and challenging. Hence, an atomic level of understanding is even more challenging, and it is highly expected to get an in-depth knowledge of the overall CRR mechanism. 
	
Cu is highly recognized as the state-of-the-art CRR material for producing multi-electron hydrocarbons \cite{sun2021achieving}, owing to advantages in terms of their abundance, non-toxic nature, better catalytic activity, high selectivity, and suitable overpotential \cite{reske2014particle,hori2008electrochemical}. To effectively enhance the activity of a catalyst, the proton-coupled electron-transfer (PCET) reactions should be effectively improved, which can be made possible by tuning the morphology or oxidation state of Cu-based nanostructures \cite{yan2021recent}. In addition to low cost, abundance, and inherent ability to form diverse nanostructures, carbon-based materials are exciting for electrocatalytic applications due to their resistance to poison and better electrochemical stability \cite{vasileff2017carbon}. Defects in carbon-based materials exist during the experimental synthesis of a large variety of materials \cite{wang2018defect,siahrostami2017theoretical} and affect the charge distribution over the surface, thereby increasing the potential active sites and improving the overall catalytic activity \cite{xue2021defect}. Defective graphene enables the formation of CO with a maximum Faraday efficiency (F.E.) of $\sim$84\% \cite{han2019defective}. The carbon-based single-atom catalysts (SAC) are highly promising for various catalytic reactions. Copper-supported defective graphene shows a strong selectivity towards methanol with an overpotential of 0.68 V\cite{he2017electrochemical}. Single Cu-atom decorated on carbon membranes catalyzes CRR to both CO and $\mathrm{CH_3OH}$ at -0.9 V \textit{vs.} RHE with an F.E. of 56\% and 44\%, respectively \cite{yang2019scalable}. 		

Biphenylene (BPH) is a planar two-dimensional (2D) allotrope of carbon (fully $\mathrm{sp^2}$ hybridized), comprised of 4-6-8 carbon rings, and has recently been experimentally synthesized. The experiment indicates metallic characteristics with an $n$-type intrinsic Dirac cone \cite{fan2021biphenylene, luo2021first}. It also displays outstanding thermal and mechanical stability, ultrahigh melting point, low thermal conductivity, and topological nature \cite{liu2021type}. It exhibits a low Gibbs free energy ($\mathrm{\Delta G_{H^*}}$) of 0.29 eV for HER, in contrast to $\mathrm{\Delta G_{H^*}}$ of 1.41 eV for graphene \cite{luo2021first}. Also, it adsorbs Li atoms are strongly compared to other carbon-containing materials and holds potential applications in Li-ion batteries \cite{denis2015hydrogen,al2022two,duhan20232,singh2023highly, sajjad2023two}. The ORR activity of BPH indicates its promising potential in alkaline fuel cells as an alternative to precious metal catalysts \cite{liu2021two}. Recently, a theoretical study\cite{lima2024unveiling} has demonstrated the adsorption of $\mathrm{CO_2}$ on pristine and defective BPH. The thermal stability and HER activity of defected-BPH are investigated in Ref.\cite{luo2023defective}, using ab-initio molecular dynamics (AIMD) simulations. The di-vacancy in BPH displays a lower value of $\mathrm{\Delta G_{H^*}}$ of -0.08 eV following the dominant Volmer-Heyrovsky mechanism for $\mathrm{H_2}$ evolution with an energy barrier of 0.80 eV. This work systematically investigated the electrocatalytic CO$_2$ reduction activity of pristine, defected, and Cu-decorated BPH using the density functional theory (DFT) calculations. The pristine BPH is inert to $\mathrm{CO_2}$ adsorption, while defected-BPH shows strong $\mathrm{CO_2}$ adsorption. Furthermore, Cu decorated BPH, a single atom catalyst (SAC), is investigated for CRR under low dispersed metal conditions. AIMD simulations are carried out to verify the clustering effects. Our results reveal the possible formation of HCOOH and methanol upon successive hydrogenation of $\mathrm{CO_2}$. A detailed mechanism of the minimum energy pathway with the rate-limiting step is discussed. Our findings shed light on BPH-based SACs for electrocatalytic CRR to valuable products and provide an adequate understanding of the catalyst selectivity regarding applied limiting potential.

\section{Computational Details}
The calculations are carried out using the projected augmented wave method, implemented in the VASP package \cite{kresse1996efficient,kresse1999ultrasoft}. The generalized gradient approximation with the Perdue-Burke-Ernzerhof scheme is used to treat the exchange-correlation functional. A plane wave basis set with an energy cutoff of 550 eV is employed to relax the 3$\times$3$\times$1 supercell, using a $\Gamma$-centered 3$\times$3$\times$1 $k$-mesh. The semi-empirical DFT-D3 dispersion correction from Grimme is utilized to account for van der Waals interactions \cite{klimevs2011van,dion2004van}. The interlayer interaction is avoided by setting a vacuum of 20 \AA\ along the perpendicular direction. The criteria for total energy and force convergence are set to $\mathrm{10^{-7}}$ eV/atom and $\mathrm{10^{-3}}$ eV/\AA, respectively. The Bader charge analysis \cite{sanville2007improved} is performed to calculate the charge distributions. The thermal stability is tested using the finite temperature AIMD simulations at 300 K. The Andersen thermostat method \cite{andersen1980molecular} is opted to control the temperature. Newton's equation of motion is integrated using the Verlet algorithm with a time step of 2 fs. The binding energy ($\mathrm{E_b}$) is defined using the formula,
\begin{equation}
\tag{1}
\mathrm{E_b} = \mathrm{E_{slab+adsorbate}} - \mathrm{E_{slab}} - \mathrm{E_{adsorbate}}
\end{equation}

where $\mathrm{E_{slab+adsorbate}}$, $\mathrm{E_{slab}}$, and $\mathrm{E_{adsorbate}}$ are the total energies of the slab with and without adsorbates, and the free energy of adsorbates (Cu as SAC, $\frac{1}{2}$$\mathrm{H_2}$, $\mathrm{CO_2}$, $\mathrm{CH_3OH}$, and $\mathrm{CH_4}$), respectively. The formation energy ($\mathrm{E_f}$) of various defects in BPH is calculated using
\begin{equation}
\tag{2}
\mathrm{E_f} = \mathrm{E_{defect}} - \mathrm{E_{pristine}} + \mathrm{n_C} \mathrm{\mu_{C}}
\end{equation}
    
where, $\mathrm{E_{defect}}$ and $\mathrm{E_{pristine}}$ indicate the total energy of the BPH sheet with and without defects, and $\mathrm{n_C}$ and $\mathrm{\mu_{C}}$ indicates the number of carbon defects created and the chemical potential of single C-atom referenced to pristine graphene, respectively. A negative value of $\mathrm{E_b}$ and $\mathrm{E_f}$ indicates an exergonic reaction and vice versa. The CRR mechanism at standard equilibrium conditions ($pH$=0, T=298.15 K and U=0) is investigated using the computational hydrogen electrode (CHE) \cite{norskov2004origin} model is employed to evaluate the Gibbs free energy ($\mathrm{\Delta G}$) of each proton-electron transfer reaction step and is defined as
\begin{equation}
\tag{3}
\mathrm{\Delta G} = \mathrm{\Delta E_b} + \mathrm{\Delta E_{ZPE}} - \mathrm{T\Delta S} + \mathrm{\Delta G_{pH}} + \mathrm{\Delta G_U}
\end{equation}

where, $\mathrm{\Delta E_b}$ is the total difference in electronic energy obtained from DFT calculations. $\mathrm{\Delta E_{ZPE}}$ is the zero point energy correction defined as equal to $\mathrm{\frac{1}{2}h\nu_i}$, where h and $\mathrm{\nu_i}$ are Planck's constant and vibrational frequency, respectively. $\mathrm{T\Delta S}$ is the vibrational entropy correction computed using phonon calculations. $\mathrm{\Delta G_{pH}}$ is the pH dependence on free energy defined as $\mathrm{k_BT\times pH\times ln(10)}$, where $\mathrm{k_B}$ is the Boltzmann constant. $\mathrm{\Delta G_U}$ (=-eU) is the influence of extra potential supplied by electrons (e) at each step. The free energy of $\mathrm{H^+}$/$\mathrm{e^-}$ pair equals $\mathrm{\frac{1}{2}\Delta G_{H_2}}$. The limiting potential ($\mathrm{U_L}$) is defined as the energy required to transform all the elementary hydrogenations to a spontaneous reaction (exergonic, negative $\mathrm{\Delta G}$) and is calculated as $\mathrm{U_L}=$$\mathrm{-\Delta G_{max}/e}$, where $\mathrm{\Delta G_{max}}$ is the maximum (positive) change in free energy. The more positive the $\mathrm{\Delta G_{max}}$, the more challenging it becomes for the elementary reactions to proceed.
  
\section{Results and Discussion}

\begin{figure}[htbp]
\centering
\includegraphics[width=\columnwidth,scale=0.6,keepaspectratio]{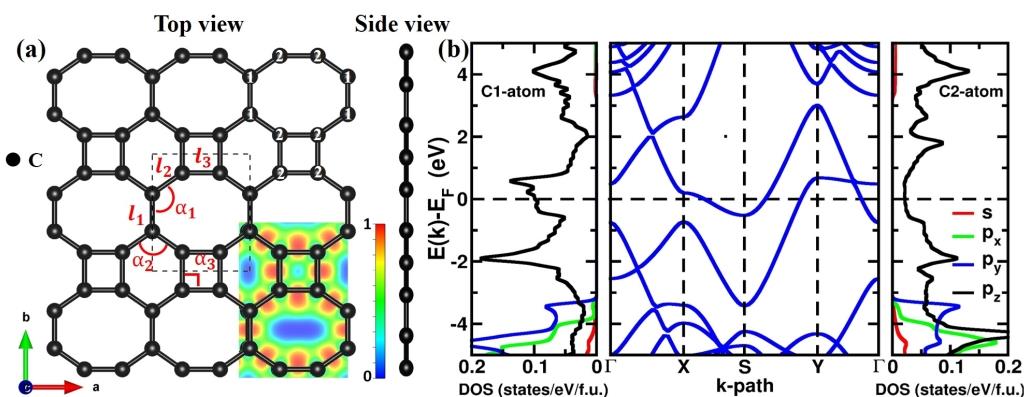}
\caption{The (a) top and side view of  BPH sheet along with electron localization function (ELF) as inset. The dotted black lines indicate a unit cell with $\alpha$ and \textit{l} representing the bond angles and bond lengths, respectively. (b) Electronic band structure and corresponding projected density of states for C1- and C2-atoms. The Fermi level is set at 0 eV.} 
\label{Fig. 1}
\end{figure}

\begin{table}[b]%
\caption{Calculated binding energies ($\mathrm{E_b}$) for all the available sites on pristine BPH when $\mathrm{CO_2}$ interacts both horizontally and vertically. h indicates the corresponding ground state height at which the molecule is adsorbed. $\mathrm{\Delta q}$ indicates the Bader charge transfer and its negative value implies charge gain by $\mathrm{CO_2}$ molecule.}
\begin{center}
\begin{ruledtabular}
\begin{tabular}{ccccccc}
& \multicolumn{6}{c}{When $\mathrm{CO_2}$ is allowed to interact} \\
\cmidrule(lr){2-7}
& \multicolumn{3}{c}{horizontal} & \multicolumn{3}{c}{vertical} \\
\cmidrule(rl){2-4} \cmidrule(rl){5-7} 
Available sites & $\mathrm{E_b}$ & h & $\mathrm{\Delta q}$ & $\mathrm{E_b}$ & h & $\mathrm{\Delta q}$ \\ [0.5ex]
				& (eV) & (\AA) & (e) & (eV) & (\AA) & (e) \\ [0.5ex]
				\hline 
				on-C1 & -0.155 & 3.28 & -0.012 & -0.086 &  3.34 & -0.005 \\ [0.5ex]
				on-C2 & -0.147 & 3.35 & -0.011 & -0.150 &  3.44 & -0.010 \\ [0.5ex]
				on-tetragonal ring & -0.140 & 3.35 & -0.009 & -0.086 &  3.18 & -0.006 \\ [0.5ex]
				on-hexagonal ring & -0.133 & 3.42 & -0.009 & -0.086 &  3.14 & -0.006 \\ [0.5ex]
				on-octagonal ring & -0.145 & 3.34 & -0.010 & -0.096 &  2.96 & -0.006 \\ [0.5ex]
			\end{tabular}
		\end{ruledtabular}
	\end{center}
	\label{table:table1}
\end{table}

The crystal structure of the Biphenylene (BPH) sheet is shown in Fig. ~\ref{Fig. 1}(a), where a primitive cell consists of six atoms arranged in a rectangular geometry with a space group Pmmm. The BPH is composed of hexagonal, tetragonal, and octagonal rings, causing a slight variation in the C$-$C bond lengths. Different atomic arrangements along the $x$- and $y$-directions suggest its anisotropic nature. The lattice constants $a$ = 3.77 \AA\ and $b$ = 4.52 \AA\ with a $b/a$ ratio of 1.19, consistent with the previous literature \cite{luo2021first,lima2024unveiling,luo2023defective}. The calculated C$-$C bond lengths \textit{l}$_1$, \textit{l}$_2$, and \textit{l}$_3$ are 1.45 \AA, 1.41 \AA, and 1.46 \AA\ and are similar to graphene with a C$-$C bond length of 1.42 \AA \. The bond angles $\alpha_1$, $\alpha_2$, and $\alpha_3$ are 124.83\textdegree, 110.33\textdegree, and 89.99\textdegree, respectively. C1 and C2 carbon atoms are inequivalent and represent the only symmetries where the density of states is specific, the rest of the atoms represent similar characteristics. As expected for sp$^2$ hybridization, all the carbon atoms are threefold coordinated \cite{liu2021type}. The electron localization function (ELF) in the inset of Fig.1(a) indicates the highest value of 0.94 in the middle of the C$-$C bond, stating the existence of strong covalent bonds between the carbon atoms. A value of 1 corresponds to perfect localization, denoted by red color. The AIMD simulations have shown that the structure remains intact at 4500 K without bond breaking but melts at 4600 K \cite{luo2021first}. The electronic band structure as in Fig. 1(b) along the $\Gamma$$-$X$-$S-Y$-$$\Gamma$ high-symmetric points indicates that BPH is metallic with an $n$-type Dirac cone observed 0.66 eV above the Fermi level ($\mathrm{E_F}$) along the $\Gamma$$-$Y direction, attributed to the out-of-plane p$_z$ orbitals of both C1- and C2-atoms, as shown in the projected density of states (PDOS). Now, we begin $\mathrm{CO_2}$ adsorption on pristine BPH considering five different possible active sites, including (1) on top of the C1-atom, (2) on the top of the C2-atom, (3) at the center of hexagonal ring (4) at the center of the tetragonal ring and (5) at the center of the octagonal ring, as in Figure S1\cite{ESI}. The linear CO$_2$ molecule is allowed to interact with these active sites horizontally and vertically (Figure S2\cite{ESI}). During optimization, the $\mathrm{CO_2}$ molecule moves away from the pristine BPH at a distance greater than 3 \AA. 
The calculated binding energy in Table~\ref{table:table1} falls in the range of -0.086 eV to -0.155 eV, indicating weak physisorption, as previously reported in Ref. \cite{lima2024unveiling}. Bader charge analysis shows a minimal charge transfer in the range of -0.005 to -0.012 e from the BPH sheet to the $\mathrm{CO_2}$ molecule, similar to graphene and graphene oxide \cite{mishra2011carbon,thakur2013ea,ma2019graphene}.

\begin{figure}[b!]
	\centering
	\includegraphics[width=\columnwidth,scale=0.6,keepaspectratio]{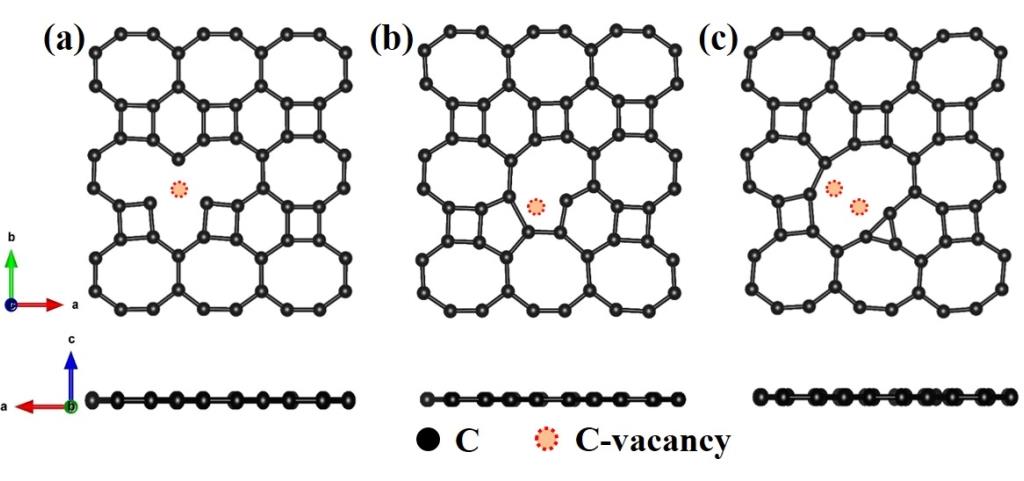}
	\caption{Optimized structures with the top and side views of various defects in BPH (a) single vacancy-C1 (SV-C1), (b) single vacancy-C2 (SV-C2), (c) double vacancy (DV)} 
	\label{Fig. 2}
\end{figure}

We next investigate the CO$_2$ adsorption in the presence of carbon vacancy in BPH, where we have considered the single vacancy (SV-C1 and SV-C2) a double vacancy (DV). Figure~\ref{Fig. 2} shows the optimized structures of SV-C1 with 4-16-4 (a 16 C-atom ring connected by two tetragons), SV-C2 with 5-9-6 (a 9 C-atom ring surrounded by a pentagon and hexagon), and DV with 4-10-3 (a 10 C-atom ring surrounded by a tetragon and triangle) carbon rings present in its skeleton. A local intrinsic magnetic moment of 1.07 $\mathrm{\mu_B}$ associated with a SV is obtained, which is attributed to unsaturated dangling bonds. Net zero magnetic moments are associated with the fully reconstructed DV, thereby saturating its dangling bonds. The calculated defect formation energy ($\mathrm{E_f}$) for SV-C1, SV-C2, and DV are -7.67 eV, -4.66 eV, and -4.97 eV, respectively. Hence, we observed that the SV-C1 defect is energetically more favorable, corroborating the previous theoretical findings \cite{luo2023defective}. These values are comparable with $\mathrm{E_f}$ of SV in graphene \cite{yang2019dft,zhang2016tight}. This suggests the possibility of SV defects in BPH, which is expected to be experimentally realized. Moreover, AIMD simulations confirm the thermal stability of various possible defects in BPH \cite{luo2023defective}.

\begin{figure}[t!]
\centering
\includegraphics[width=\columnwidth,scale=0.6,keepaspectratio]{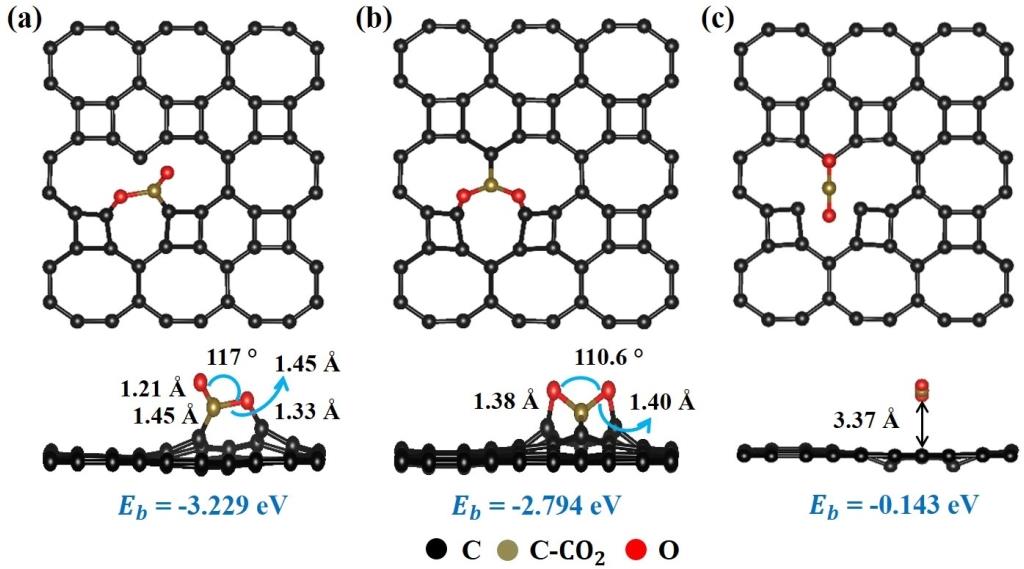}
\caption{Optimized structures with the top and side views of $\mathrm{CO_2}$ adsorption on SV-C1 defect when $\mathrm{CO_2}$ is placed (a) horizontally with O-atom of $\mathrm{CO_2}$ facing the vacancy site, (b) horizontally with C-atom of $\mathrm{CO_2}$ facing the vacancy site, and (c) vertically at the top of vacancy site.} 
\label{Fig. 3}
\end{figure}

The $\mathrm{CO_2}$ molecule can interact horizontally and vertically with these vacancy sites. The defects strengthen the $\mathrm{CO_2}$ adsorption energy, as evident from its proximity with the defective BPH sheet. The optimized configurations for the $\mathrm{CO_2}$ adsorption at all the vacancy sites are depicted in Figure S3 \cite{ESI}. When the $\mathrm{CO_2}$ molecule is placed vertically, it reorients itself and prefers to align horizontally over the vacancy site. And when placed horizontally, only SV-C1 defect is found to activate $\mathrm{CO_2}$ molecule depicting strong binding energy. Except for SV-C1, for all other defect configurations, the $\mathrm{CO_2}$ molecule goes away by more than 2.85 \AA\, indicating weak physisorption energy in the range of 0.262 eV to -0.178 eV. This can also be inferred from the Bader charge analysis, which indicates infinitely small charge transfer from the sheet to $\mathrm{CO_2}$ molecule in the range of 0.013 to 0.018 e. We conclude that only SV-C1 defect is favorable for $\mathrm{CO_2}$ adsorption when aligned horizontally. To better confirm the $\mathrm{CO_2}$ interactions on SV-C1, it is placed horizontally in two different ways. One where the O atom of $\mathrm{CO_2}$ is facing the vacancy site and the other where the C atom of $\mathrm{CO_2}$ is facing the vacancy site. The former indicates a $\mathrm{E_b}$ of -3.23 eV (henceforth labeled as SV1), whereas the latter indicates a $\mathrm{E_b}$ of -2.79 eV (henceforth labeled as SV2). In both cases, a strong atomic arrangement disrupts the planar-defected sheet locally. The O-C-O bond angle of the linear $\mathrm{CO_2}$ molecule reduces to 117.0\textdegree\, and both the O-C bond lengths drastically increase from 1.18 \AA\ to 1.21 \AA\ and 1.45 \AA, thus bringing $\mathrm{CO_2}$ molecule near the defected area. For SV2, the O-C-O angle of the linear $\mathrm{CO_2}$ molecule reduces to 110.6\textdegree\, and the O-C bond lengths increase from 1.18 \AA\ to 1.40 \AA. Meanwhile, the presence of strong interactions between $\mathrm{CO_2}$ and SV1 (SV2) is also evident from the Bader charge analysis indicating a charge transfer of 0.759 e (1.262 e) from the defected BPH to the $\mathrm{CO_2}$ molecule. No intrinsic magnetic moments are observed for horizontally aligned $\mathrm{CO_2}$ molecule. Since SV1 indicates a much more vital binding energy, we investigate it for CRR. The vertical configuration does not bind $\mathrm{CO_2}$, indicating weak physisorption energy with a $\mathrm{E_b}$ of -0.143 eV. The molecule goes away by 3.37 \AA. This can be inferred from an infinitely small charge transfer of 0.016 e from the sheet to the $\mathrm{CO_2}$ molecule, as evident from the Bader charge analysis.

By using a single electron PCET mechanism, we will examine the formation of all the $\mathrm{C_1}$ value-added hydrocarbon products (CO, HCOOH, $\mathrm{CH_3OH}$, and $\mathrm{CH_4}$). A schematic representation of the possible elementary reaction pathways for the electrochemical CRR on BPH is depicted in Figure S4 \cite{ESI}. The Gibbs free energy ($\mathrm{\Delta G}$) is an essential catalytic descriptor of all the various reaction intermediates formed during the PCET mechanism at 300 K. Here * indicates the point of interaction of $\mathrm{CO_2}$ and its intermediates with the defected BPH. However in our case, $\mathrm{CO_2}$ prefers to bond strongly with SV1 by forming $\mathrm{C_{CO_2}-C_{BPH}}$ and $\mathrm{O_{CO_2}-C_{BPH}}$ bond with a bond lengths of 1.45 \AA\ and 1.33 \AA, respectively. Our calculations suggest that H prefers to bind with O-atom of $\mathrm{CO_2}$ and reduces adsorbed $\mathrm{CO_2}$ to *COOH with a $\mathrm{E_b}$ of -1.85 eV, indicating exergonic reaction. The generated *COOH intermediate gets further attacked by incoming $\mathrm{H^+}$/$\mathrm{e^-}$. If the O-atom of O.H. present in *COOH gets attacked by $\mathrm{H^+}$/$\mathrm{e^-}$, then there is a possibility of C.O. formation with the removal of a water molecule. In contrast, if C-atom gets attacked, then HCOOH can be formed. Such a reaction is commonly observed in metal, metal oxides, or metal chalcogenides \cite{kovacic2020photocatalytic}. Interestingly, in our case, the C-atom of $\mathrm{CO_2}$ gets attracted more towards the vacancy site, and three C$-$O bonds exist (two $\mathrm{C_{CO_2}-C_{BPH}}$ bonds and one $\mathrm{O_{CO_2}-C_{BPH}}$ bond). So the OH present in *COOH is more favorable compared to $\mathrm{C_{CO_2}}$ and $\mathrm{O_{CO_2}}$ atoms in *COOH. Hence, $\mathrm{H^+}$/$\mathrm{e^-}$ attacks O.H. of *COOH and helps remove a water molecule leaving *C.O. At this point, we observe that $\mathrm{C_{CO_2}}$ gets attracted towards the vacancy site and tries to occupy it. We can also see the formation of two C-O bonds, a $\mathrm{C_{CO_2}-C_{BPH}}$ bond and another $\mathrm{O_{CO_2}-C_{BPH}}$ bond of which C-atom of $\mathrm{CO_2}$ now becomes a part of BPH structure. This is an attempt to saturate the dangling bonds at the vacancy site. Our calculations indicate a reduction of *COOH to *C.O. with a reduction of a water molecule, indicating a $\mathrm{E_b}$ of -1.483 eV. As a consequence of this, the O-atom of $\mathrm{CO_2}$ lies above the sheet and remains the only possibility of the next $\mathrm{H^+}$/$\mathrm{e^-}$ attack. If now a $\mathrm{H^+}$/$\mathrm{e^-}$ attacks O-atom of $\mathrm{CO_2}$, then *COH intermediate can be formed. Our calculations suggest that the H-atom prefers to interact with the O-atom of *C.O. and is further reduced to *COH, indicating a $\mathrm{E_b}$ of -1.682 eV. Finally, another water molecule comes out with the last $\mathrm{H^+}$/$\mathrm{e^-}$ attack on the O.H. group of *COH. Our calculations suggest that as soon as an H-atom interacts with the O.H. group of *COH, a water molecule is formed, turning the SV-C1 defected sheet back to defect-free BPH, indicating a $\mathrm{E_b}$ of -0.875 eV. This indicates the inertness of BPH towards $\mathrm{CO_2}$ reduction. We have also investigated the electrocatalytic CRR on SV2 to verify any possibility for $\mathrm{CO_2}$ reduction. This configuration follows a similar reduction cycle as SV1, where *$\mathrm{CO_2}$ is first reduced to *COOH, indicating a $\mathrm{E_b}$ of -2.794 eV. With the next $\mathrm{H^+}$/$\mathrm{e^-}$ attack on the O.H. group of *COOH, *C.O. is formed and follows a similar trend of intermediates and $\mathrm{E_b}$. All the binding energies are negative, indicating an exergonic reaction. Hence, we observe that the presence of vacancy defects in BPH is ineffective in catalyzing CRR and, therefore, is not an effective strategy to alter its electronic properties.

We now investigate the Cu-decorated pristine BPH as a SAC because Cu has numerous advantages, including its abundance, inexpensive, environment friendly, and highly active in catalyzing $\mathrm{CO_2}$ to methanol \cite{shen2018cu,cometto2021copper}. The Cu-atom is placed at various available active sites on the surface of BPH and can be optimized to obtain minimum energy configuration (see Fig. S6 in Supporting Information\cite{ESI}). The calculated $\mathrm{E_b}$ when Cu-atom is placed on the top of C1-atom, on the top of C2-atom, at the center of the tetragonal ring, at the center of the hexagonal ring, and the center of the octagonal ring are -0.873, -1.28 eV, -1.06 eV, -0.75 eV, and -1.28 eV, respectively. Among all, the Cu-atom prefers to adsorb on top of the C2-atom at a distance of 1.91 \AA\, indicating a strong $\mathrm{E_b}$ of -1.28 eV. Bader charge analysis reveals that Cu-atom loses a charge of 0.225 e to the BPH, depicting strong $\mathrm{C_{BPH}-Cu}$ covalent bond formation. Cu-atoms should not form clusters to utilize Cu-BPH SACs for CRR applications effectively. In our case, the most favorable configuration for Cu-BPH SAC has a $\mathrm{E_b}$ of -1.28 eV, which is higher than the bulk cohesive energy ($\mathrm{E_{coh}}$) energy (-3.49 eV). If two Cu-atoms are at a distance much greater than their bulk bond length of 2.35 \AA\, they will not tend to agglomerate. To establish this, we have performed the AIMD calculations at 300 K to investigate the possibility of clustering (see Fig. S7 in Supporting Information\cite{ESI}). It is observed that when Cu atoms are placed at a distance of $\sim$3.38 \AA\, they show a tendency to form a Cu-Cu cluster with a bond length of 2.35 \AA\, while when they are kept far apart at a double distance of $\sim$8 \AA\, challenging to form a metal cluster. Such a situation can be achieved experimentally by dispersing metal atoms in minimal concentrations in a controlled manner.

\begin{figure}[t!]
\centering
\includegraphics[height=3.5cm, width=\columnwidth]{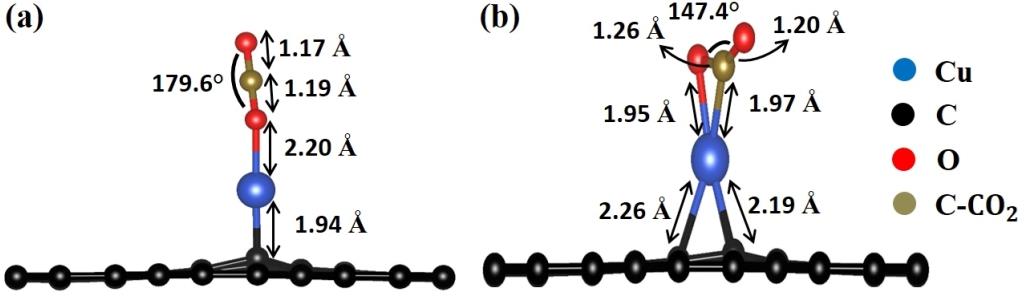}
\caption{Optimized structures of $\mathrm{CO_2}$ molecule adsorbed over Cu-decorated pristine BPH when $\mathrm{CO_2}$ is placed (a) vertically and (b) horizontally at its top.} 
\label{Fig. 4}
\end{figure}

Figure~\ref{Fig. 4} indicates the optimized structures of $\mathrm{CO_2}$ molecule on Cu-BPH. The $\mathrm{CO_2}$ molecule is adsorbed at a distance of 2.20 \AA\ from the Cu-BPH when placed vertically, indicating weak adsorption energy with a $\mathrm{E_b}$ of -0.34 eV, indicating its inability to activate $\mathrm{CO_2}$. In contrast, the $\mathrm{CO_2}$ molecule activates favorably when aligned horizontally over Cu-BPH. It is interesting to observe that the adsorption of $\mathrm{CO_2}$ aligns Cu on the top of a square carbon ring, making four $\mathrm{C2_{BPH}-Cu}$ bonds with bond lengths 2.19, 2.36, 2.25, and 1.98 \AA. The linear bond angle of $\mathrm{CO_2}$ decreases to 147.4\textdegree\ with an increase in the bond lengths from 1.18 \AA\ to 1.26 \AA\ and 1.20 \AA, respectively. The C- and O-atoms of $\mathrm{CO_2}$ are bonded with the Cu-atom making $\mathrm{C_{CO_2}-Cu}$ and $\mathrm{O_{CO_2}-Cu}$ bonds having bond lengths 1.97 \AA\ and 1.95 \AA, respectively. Compared to the vertical configuration, this configuration yields stronger adsorption with a $\mathrm{E_b}$ of -0.52 eV. It is also evident from the Bader charge analysis that $\mathrm{CO_2}$ molecule gains a charge of 0.39 e from the Cu-BPH SAC. We therefore investigate it further for the electrocatalytic reduction of $\mathrm{CO_2}$ to HCOOH and $\mathrm{CH_3OH}$.

\begin{figure*}[t!]
\centering
\includegraphics[height=12cm, width=15cm]{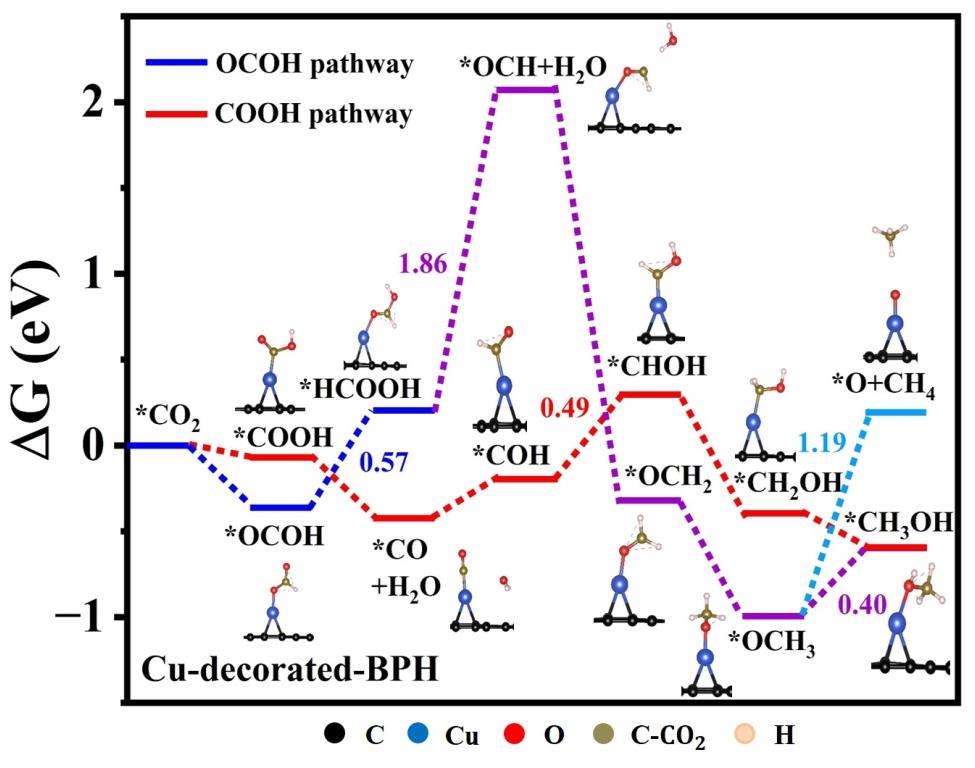}
\caption{Gibbs free energy ($\Delta$G) profile for $\mathrm{CO_2}$ reduction over Cu-BPH SAC. The inset figure indicates the corresponding optimized geometries for the reaction intermediates during the $\mathrm{CO_2}$ reduction process. The color scheme represents the formation of different product hydrocarbons. Blue color lines indicate the 2$\mathrm{e^-}$ reduction reaction of adsorbed $\mathrm{CO_2}$ to HCOOH, red color lines indicate the 6$\mathrm{e^-}$ reduction reaction of adsorbed $\mathrm{CO_2}$ to $\mathrm{CH_3OH}$, purple color lines indicate an unfavorable reaction forming methanol and methane (sky blue color line).} 
\label{Fig. 5}
\end{figure*}	
	
The CRR mechanism is highly explored on transition metal surfaces\cite{siahrostami2017theoretical} using the CHE model proposed by Norskov and coworkers\cite{norskov2004origin}. At pH=0 and T=300 K, we have investigated various possibilities of forming $\mathrm{C_1}$-value-added hydrocarbons. A thermo-chemical $\mathrm{CO_2}$ reduction pathway involving the formation of 2$\mathrm{e^-}$ (HCOOH, CO) and 6$\mathrm{e^-}$ ($\mathrm{CH_3OH}$) hydrocarbon products undergoes through formation of various intermediates. The initial hydrogenation of adsorbed $\mathrm{CO_2}$ molecule either forms $\mathrm{^*{COOH}}$/$\mathrm{^*{OCOH}}$, which on second hydrogenation can be reduced to $\mathrm{^*{CO}}$/$\mathrm{^*{HCOOH}}$ (see Eqn. 4 and 5). So, the 2$\mathrm{e^-}$ reduction products can be further desorbed from the catalyst surface as CO or HCOOH.

\begin{equation}
\mathrm{CO_2} \rightarrow \mathrm{^*{COOH}} \rightarrow \mathrm{^*{CO}} \rightarrow \mathrm{CO}
\tag{4}
\end{equation}

\begin{equation}
\mathrm{CO_2} \rightarrow \mathrm{^*{OCOH}}  \rightarrow \mathrm{^*{HCOOH}} \rightarrow \mathrm{HCOOH}
\tag{5}
\end{equation}
	
If the adsorbed $\mathrm{^*{CO}}$ molecule has a stronger binding then it gets further reduced to $\mathrm{^*{COH}}$ forming $\mathrm{CH_3OH}$ as product (see Eqn. 6). Similarly, if adsorbed $\mathrm{^*{HCOOH}}$ intermediate gets further reduced to $\mathrm{^*{OCH}}$ or $\mathrm{^*{COH}}$, then $\mathrm{CH_3OH}$ and $\mathrm{CH_4}$ are the available hydrocarbon products formed after successive six and eight hydrogenations.
	
\begin{equation}
\mathrm{^*{CO}} \rightarrow \mathrm{^*{COH}} \rightarrow \mathrm{^*{CHOH}} \rightarrow \mathrm{^*{CH_2OH}} \rightarrow \mathrm{^*{CH_3OH}}
\tag{6}
\end{equation}
	
\begin{multline}
\mathrm{^*{HCOOH}} \rightarrow \mathrm{^*{OCH+H_2O}} \rightarrow \mathrm{^*{OCH_2}} \rightarrow \mathrm{^*{OCH_3}} \\
\rightarrow \mathrm{^*{O+CH_4}} \rightarrow \mathrm{^*{OH}} \rightarrow \mathrm{H_2O}
\tag{7}
\end{multline}
	
However, there remains a possibility of the formation of $\mathrm{CH_4}$ as the product hydrocarbon if $\mathrm{^*{OCH_3}}$ gets further reduced (see Equ. 7). The proposed schematic representation of these possible pathways is shown in Fig. S4 in Supporting Information \cite{ESI}.

	\begin{table}[b]%
	\caption{Comparison of the rate-limiting potentials ($\mathrm{U_L}$) of various studied catalysts for the production of HCOOH and $\mathrm{CH_3OH}$.}
	\begin{center}
		\begin{ruledtabular}
			\begin{tabular}{cccc}
				Catalyst & $\mathrm{U_L}$ (eV) & Production & Reference \\ [0.5ex] \hline
				Cu/m-VacG & 1.18 & HCOOH & \cite{ali2021stability} \\ [0.5ex]
				CuPc/graphene & 0.95  & HCOOH &  \\ [0.5ex]
				& 1.41 & $\mathrm{CH_3OH}$ & \cite{yan2022theoretical} \\ [0.5ex]
				Cu@$\mathrm{N_b}$-doped GDY & 0.65  & HCOOH & \cite{feng2021theoretical} \\ [0.5ex]
				$\mathrm{Co_3}$@$\mathrm{C_2N}$ & 0.71  & HCOOH & \cite{zha2021efficient} \\ [0.5ex]
				Cu@$\mathrm{C_2N}$ & 1.81  & $\mathrm{CH_3OH}$ & \cite{cui2018c} \\ [0.5ex]		
				Cu@dv-Gr & 1.48  & $\mathrm{CH_3OH}$ & \cite{back2017single} \\ [0.5ex]
				octahedral  $\mathrm{Cu_{85}}$ \\ nanocluster & 0.53 & $\mathrm{CH_3OH}$ & \cite{rawat2017thermochemical} \\ [0.5ex]
				$\mathrm{Cu_3}$/ND@Gr & 0.58-0.70 & $\mathrm{CH_3OH}$ & \cite{du2022anchoring} \\ [0.5ex]
				Cu on single vacancy \\ graphene & 0.65 & $\mathrm{CH_3OH}$ & \cite{he2017electrochemical} \\ [0.5ex]
				Cu(111) ML & 0.46 & $\mathrm{CH_3OH}$ & \cite{mandal2019hexagonal} \\ [0.5ex]
				Si-doped graphene edges & 0.49 & $\mathrm{CH_3OH}$ & \cite{mao2019silicon} \\ [0.5ex]
				Cu-BPH & 0.49 & $\mathrm{CH_3OH}$ & Our study \\ [0.5ex]
				Cu-BPH & 0.57 & HCOOH & Our study \\ [0.5ex]
			\end{tabular}
		\end{ruledtabular}
	\end{center}
	\label{table:table2}
\end{table}	 
	
Figure~\ref{Fig. 5} depicts the Gibbs free energy ($\mathrm{\Delta G}$) profile for the $\mathrm{CO_2}$ reduction on Cu-BPH SAC indicating two different pathways with the formation of possible products and their corresponding limiting potentials ($\mathrm{U_L}$). It is clear that the first $\mathrm{H^+}$/$\mathrm{e^-}$ attack on adsorbed $\mathrm{CO_2}$ forms formate group ($\mathrm{^*{OCOH}}$) more favorably compared to the carboxyl group ($\mathrm{^*{COOH}}$). The $\mathrm{^*{OCOH}}$ ($\mathrm{^*{COOH}}$) intermediate is formed when an H-atom interacts with C (O)-atom of $\mathrm{CO_2}$ indicating a $\mathrm{E_b}$ of -0.669 (-0.365) eV with an adsorption height of 1.813 (1.909) \ A.A. The generated $\mathrm{^*{OCOH}}$/$\mathrm{^*{COOH}}$ undergoes further reduction with next $\mathrm{H^+}$/$\mathrm{e^-}$. The second $\mathrm{H^+}$/$\mathrm{e^-}$ attack on O-atom present in $\mathrm{^*{OCOH}}$ forms *HCOOH with a $\mathrm{E_b}$ of 0.029 eV, suggesting small endergonic reaction. The bond length between Cu-BPH and O-atom present in $\mathrm{^*{HCOOH}}$ increases to 1.899 \ A.A. \ from 1.813 \ A.A. \, suggesting that they are loosely bound. In other words, an increased bond length and small value of $\mathrm{E_b}$ hints towards the possible evolution of HCOOH. In such a case, the conversion from $\mathrm{^*{OCOH}}$ to $\mathrm{^*{HCOOH}}$ will be the rate-limiting step with a $\mathrm{U_L}$ of 0.57 eV.

In contrast, we also tried to investigate any possibility of further $\mathrm{H^+}$/$\mathrm{e^-}$ attack on $\mathrm{^*{HCOOH}}$. With the next $\mathrm{H^+}$/$\mathrm{e^-}$ attack, the $\mathrm{^*{HCOOH}}$ is reduced to $\mathrm{^*{OCH+H_2O}}$ with a $\mathrm{E_b}$ of 1.787 eV, suggesting highly endergonic reaction. This reduction step will now become the rate-limiting step for the entire reduction process with a $\mathrm{U_L}$ of 1.87 eV. This on successive $\mathrm{H^+}$/$\mathrm{e^-}$ attacks forms methanol and methane with a second $\mathrm{U_L}$ of 0.40 eV and 1.19 eV, respectively. It can be clearly understood that the first $\mathrm{U_L}$ of 1.87 eV is itself sufficiently large enough. 

The requirement of such a large free energy value indicates a highly unfavorable reduction mechanism. Thus indicating a higher possibility for the evolution of HCOOH as product hydrocarbon with 2$\mathrm{e^-}$ reduction mechanism. Now, if the $\mathrm{H^+}$/$\mathrm{e^-}$ attacks on C-atom present in $\mathrm{^*{COOH}}$ then $\mathrm{^*{CO}}$ and a water molecule is formed with a $\mathrm{E_b}$ of -0.614 eV. The increase in the value of binding energy can be inferred from the decrease in bond length between C-atom present in $\mathrm{^*{C.O.}}$ and Cu-BPH, which decreases to 1.779 \ A.A. \ from 1.909 \ A.A. \ suggesting strong interaction. With the third $\mathrm{H^+}$/$\mathrm{e^-}$, the H-atom interacts with the C-atom present on $\mathrm{^*{CO}}$ and forms $\mathrm{^*{COH}}$ with a $\mathrm{E_b}$ of 0.887 eV. An increase in the bond length is observed between C-atom (present in $\mathrm{^*{COH}}$) and Cu-BPH from 1.779 \ A.A. to 1.911 \ A.A. With the next $\mathrm{H^+}$/$\mathrm{e^-}$, $\mathrm{^*{COH}}$ is reduced to $\mathrm{^*{CHOH}}$ as H-atom finds more favorable to bind with O-atom giving a $\mathrm{E_b}$ of -0.098 eV indicating a bond length of 1.816 \ A.A. This is one fifth $\mathrm{H^+}$/$\mathrm{e^-}$, reduces to $\mathrm{^*{CH_2OH}}$ with H-atom prefer to bind with C-atom (present in $\mathrm{^*{CH_2OH}}$) indicating a $\mathrm{E_b}$ of -0.789 eV and an increase in Cu-BPH and C-atom (present in $\mathrm{^*{CH_2OH}}$) bond length of 1.928 \ A.A. Finally with the last $\mathrm{H^+}$/$\mathrm{e^-}$, H-atom binds favorably with the C-atom of $\mathrm{^*{CH_2OH}}$ forming $\mathrm{^*{CH_3OH}}$ with a binding energy of -0.792 eV. The bond length between Cu-BPH and C-atom (present in $\mathrm{^*{CH_3OH}}$) increases to 1.944 \ A.A. Clearly, the reduction of $\mathrm{^*{HCOO}}$, $\mathrm{^*{C.O.}}$, $\mathrm{^*{COH}}$ is endergonic, while the remaining reactions are exergonic. Hence for the evolution of HCOOH, the conversion of $\mathrm{^*{HCOO}}$ to $\mathrm{^*{HCOOH}}$ is the rate-limiting step with a $\mathrm{U_L}$ of 0.57 eV. Whereas for the formation of $\mathrm{^*{CH_3OH}}$, the conversion of $\mathrm{^*{COH}}$ to $\mathrm{^*{CHOH}}$ is the rate-limiting step with a $\mathrm{U_L}$ of 0.49 eV. Our calculated results for $\mathrm{U_L}$ are found to be much lower compared to the well-known Cu electrodes requiring an overpotential of $\sim$1 V. Additionally, a comparison of limiting potentials for HCOOH and $\mathrm{CH_3OH}$ as in Table~\ref{table:table2} indicates that our calculated results are much smaller or comparable with other studied catalysts.

\begin{figure}[b!]
	\centering
	\includegraphics[width=\columnwidth,scale=0.6,keepaspectratio]{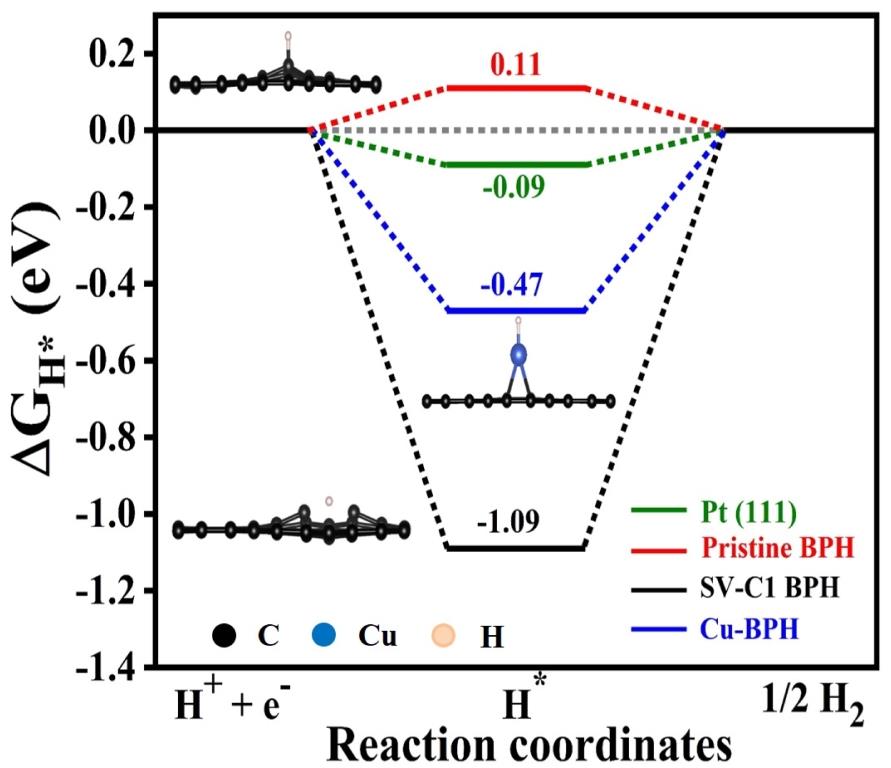}
	\caption{Gibbs free energy profile for HER on perfect, SV-C1 BPH, and Cu-BPH SAC, respectively. The inset figure represents the corresponding minimum energy configurations. Pt(111) is shown for a reference comparison, taken from \cite{norskov2005trends}.} 
	\label{Fig. 6}
\end{figure}
		
One competing reaction during the $\mathrm{CO_2}$ reduction cycle is the hydrogen evolution reaction (HER). We investigated the HER activity of pristine, SV-C1 BPH, and Cu-BPH SAC as in Fig.~\ref{Fig. 6}. Our calculated results reveal that H-atom prefers to interact with C2-atom with a $\mathrm{\Delta G_{H^*}}$ of 0.11 eV, which is consistent with previous study\cite{luo2021first}. For SV-C2 BPH, the H-atom interacts at the vacancy site with a $\mathrm{\Delta G_{H^*}}$ of -1.17 eV. For Cu-BPH SAC, the H-atom interacts with the Cu-atom at the height of 1.505 \AA\, giving a $\mathrm{\Delta G_{H^*}}$ of -0.47 eV. Ideally, a small value of $\mathrm{\Delta G_{H^*}}$ close to zero indicates a more favorable HER side reaction which will dominate the CRR. Lastly, The electrocatalytic CRR mechanism in Cu-BPH catalyst is investigated using the charged hydrogen pseudopotential in our simulations (see Figure S8 \cite{ESI}). The first protonation form $\mathrm{^*COOH}$ with a strong $\mathrm{E_b}$ of -1.19 eV. The second protonation reduces $\mathrm{^*COOH}$ to $\mathrm{^*CO}$, and a water molecule is released, indicating a stronger $\mathrm{E_b}$ of -1.205 eV. The enhanced value of binding energy is attributed to the C$-$O bond breaking together with the removal of a $\mathrm{H_2O}$ molecule. It is also observed that the bond length between C$-$ (O$-$) atom (present in $\mathrm{^*CO}$) with Cu atom reduces (increases) slightly to 1.89 (2.03) \AA\ from 1.92 (1.97) \AA. In third protonation, $\mathrm{^*COH}$ is formed with a $\mathrm{E_b}$ of -0.956 eV. $\mathrm{^*CH_2O}$ is formed in the next protonation with $\mathrm{E_b}$ of -0.829 eV and a small increase in the bond length of Cu$-$O (present in $\mathrm{^*CH_2O}$). Interestingly with next protonation, $\mathrm{^*CH_2O}$ is reduced to $\mathrm{^*CH_2OH}$ instead of $\mathrm{^*CH_3O}$ indicating a $\mathrm{E_b}$ of -0.824 eV and a small increase in the bond length of 2.05 \AA\ between the O atom (present in $\mathrm{^*CH_2OH}$) and Cu-atom. In this step, protons readily bind with the O atom as the C atom has already shared its four valence electrons. Finally, in the last protonation, $\mathrm{^*CH_3OH}$ is formed with a $\mathrm{E_b}$ of -0.79 eV with the catalyst. All the binding energy values are negative, indicating an exergonic reaction. We see that the protonation of adsorbed $\mathrm{CO_2}$ on Cu-BPH SAC follows $\mathrm{^*CO_2}$ $\rightarrow$ $\mathrm{^*COOH}$ $\rightarrow$ $\mathrm{^*CO+H_2O}$ $\rightarrow$ $\mathrm{^*COH}$ $\rightarrow$ $\mathrm{^*CH_2O}$ $\rightarrow$ $\mathrm{^*CH_2OH}$ $\rightarrow$ $\mathrm{^*CH_3OH}$ and thus indicates possible formation of methanol as product. We notice a continuous increase in the bond lengths between the reaction intermediates and the Cu-BPH, which hints towards the possible evolution of methanol as a product hydrocarbon.
	
\section{Conclusions}
In summary, we have explored the $\mathrm{CO_2}$ catalytic activity of pristine, defective, and Cu decorated Biphenylene (Cu-BPH). The pristine BPH is chemically inert towards $\mathrm{CO_2}$ molecule. A single carbon vacancy strongly attracts the C-atom of $\mathrm{CO_2}$ towards the vacancy site. The Cu decorated BPH can also activate $\mathrm{CO_2}$ molecule with a binding energy ($\mathrm{E_b}$) of -0.517 eV and with a charge transfer of 0.391 e from the Cu-BPH to the $\mathrm{CO_2}$ molecule. The kinetics reveal that the Cu-atom avoids clustering, as evidenced by the molecular dynamics calculations. The calculated Gibbs free energy ($\mathrm{\Delta G}$) profile reveals that the formate pathway generates $\mathrm{^*HCOOH}$ with the conversion of $\mathrm{^*OCOH}$ to $\mathrm{^*HCOOH}$ as the rate-limiting potential ($\mathrm{U_L}$) of 0.57 eV. After the 2$\mathrm{e^-}$ reduction step, there is a possibility of forming formic acid as indicated by a small $\mathrm{E_b}$ of 0.029 eV. In the case of $\mathrm{^*COOH}$ pathway, the $\mathrm{\Delta G}$ confirms the conversion of $\mathrm{^*COH}$ to $\mathrm{^*CHOH}$ as the rate-limiting step for the formation of $\mathrm{CH_3OH}$ with a $\mathrm{U_L}$ of 0.49 eV. The protonation model also confirms the possible formation of $\mathrm{CH_3OH}$. Our calculated $\mathrm{U_L}$ is lower than the commercial Cu electrodes and thus shows promise for electrocatalytic CRR. We believe our work will provide useful insights into understanding the catalytic activity of carbon-based SACs.
	
\section*{Acknowledgements}
R.S. acknowledges IIT Bombay for providing Institute postdoctoral fellowship and computational resources to pursue this work. N.S. acknowledges the financial support from Khalifa University of Science and Technology under the Research Innovation Grant (RIG)-2023-01. 

\bibliography{references.bib} 
\end{document}


\title{Supplimentary Information\\ Efficient Electrochemical CO$_2$ Reduction Reaction over Cu-decorated Biphenylene}

\author{Radha N Somaiya}
\affiliation{Materials Modeling Laboratory, Department of Physics, IIT Bombay, Powai, Mumbai 400076, India}

\author{Muhammad Sajjad}
\affiliation{Nottingham Ningbo China Beacons of Excellence Research and Innovation Institute, University of Nottingham, Ningbo, China}
\affiliation{Key Laboratory of Carbonaceous Wastes Processing and Process Intensification Research of Zhejiang Province, University of Nottingham Ningbo China, Ningbo 315100, China}
		
\author{Nirpendra Singh}
\email{Nirpendra.Singh@ku.ac.ae}
\affiliation{Department of Physics, Khalifa University, Abu Dhabi, 127788, United Arab Emirates}
\affiliation{Research and Innovation Center for Graphene and 2D materials (RIC2D), Khalifa University, Abu Dhabi, United Arab Emirates}

\author{Aftab Alam}
\email{aftab@iitb.ac.in}
\affiliation{Materials Modeling Laboratory, Department of Physics, IIT Bombay, Powai, Mumbai 400076, India}
\maketitle

Here, we have provided additional information about the possible adsorption sites on the surface of Biphenylene (BPH), optimized structures of $\mathrm{CO_2}$ adsorbed on pristine and defected BPH, a proposed schematic electrochemical $\mathrm{CO_2}$ reduction pathway, $\mathrm{CO_2}$ reduction reaction on single vacancy-C1 BPH, optimized configurations of Cu-atom decorated on pristine BPH, molecular dynamics simulations for Cu-BPH SAC, $\mathrm{CO_2}$ reduction reaction on Cu-BPH under the effect of proton to strengthen our message in the manuscript.

\begin{figure*}[b]
	\centering
	\includegraphics[width=0.8\columnwidth,keepaspectratio]{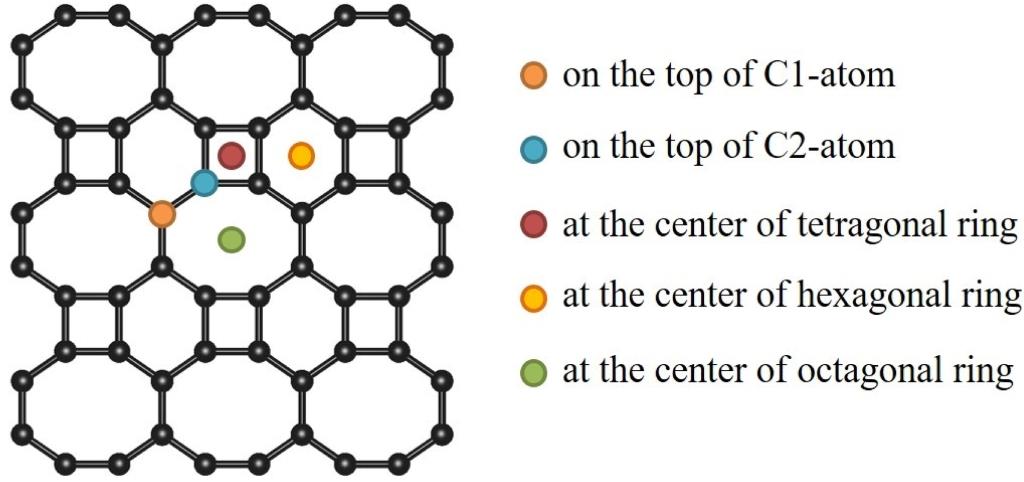}
	\caption{Possible adsorption sites on the surface of BPH sheet. C and H atoms are indicated by black and pink spheres, respectively} 
	\label{Fig. S1}
\end{figure*}

\begin{figure*}[t]
	\centering
	\includegraphics[width=\columnwidth,keepaspectratio]{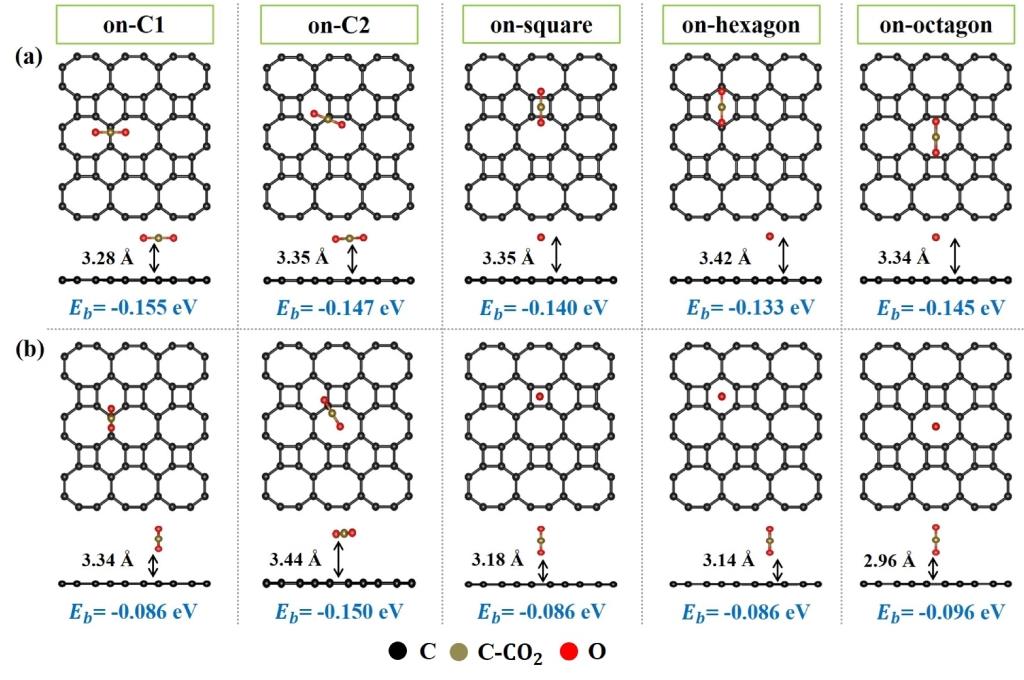}
	\caption{Optimized geometries of $\mathrm{CO_2}$ molecule adsorbed over pristine Biphenylene network (BPH) when $\mathrm{CO_2}$ molecule is allowed to interact (a) horizontally and (b) vertically over the five different available active sites. $\mathrm{E_b}$ represents the corresponding binding energies. The distance between the molecule and BPH is indicated with an arrow.} 
	\label{Fig. S2}
\end{figure*}

\begin{figure*}[t]
	\centering
	\includegraphics[width=\columnwidth,keepaspectratio]{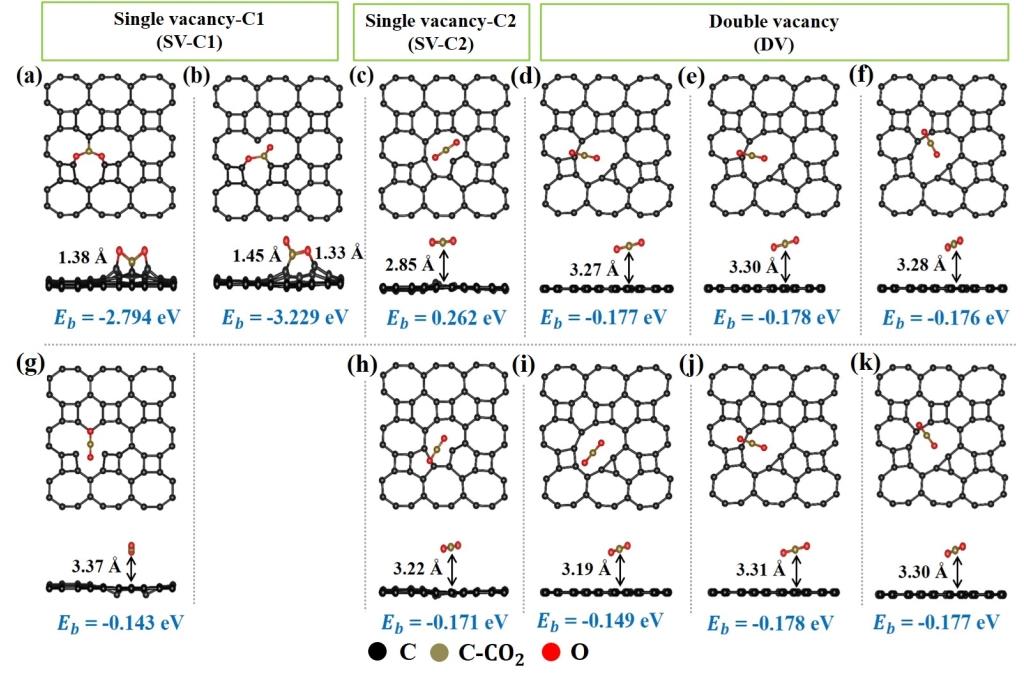}
	\caption{Optimized geometries of $\mathrm{CO_2}$ molecule adsorbed over defected Biphenylene network (BPH) when the $\mathrm{CO_2}$ molecule is placed (a-f) horizontally and (g-k) vertically over the single vacancy-C1 (SV-C1), single vacancy-C2 (SV-C2), and double vacancy (DV) sites. At all the vacancy sites, the $\mathrm{CO_2}$ molecule was placed differently to find the minimum ground state configuration.} 
	\label{Fig. S3}
\end{figure*}

\begin{figure*}[b]
	\centering
	\includegraphics[width=\columnwidth,keepaspectratio]{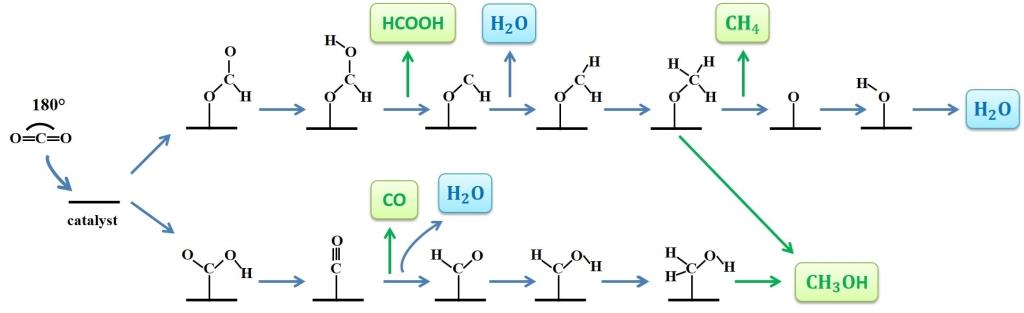}
	\caption{Proposed schematic representation of the reaction pathways for the electrochemical reduction of $\mathrm{CO_2}$ to product hydrocarbons (CO, HCOOH, $\mathrm{CH_4}$, and $\mathrm{CH_3OH}$) on Biphenylene.} 
	\label{Fig. S4}
\end{figure*}

\begin{figure*}[t]
	\centering
	\includegraphics[width=\columnwidth,keepaspectratio]{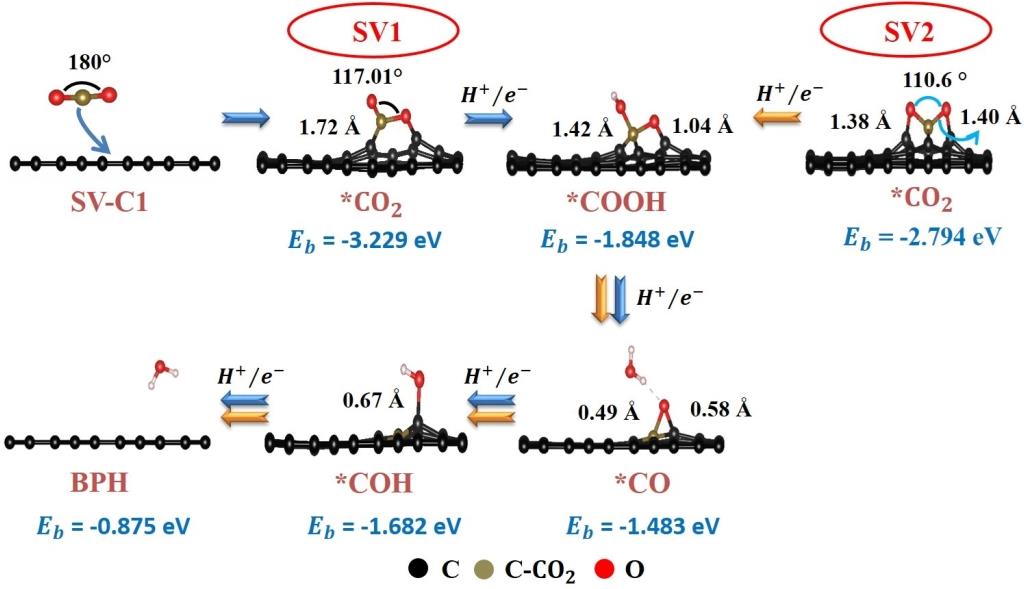}
	\caption{The electrocatalytic $\mathrm{CO_2}$ reduction reaction mechanism on Single vacancy-C1 (SV-C1) BPH. } 
	\label{Fig. S5}
\end{figure*}

\begin{figure*}[b]
	\centering
	\includegraphics[width=\columnwidth,keepaspectratio]{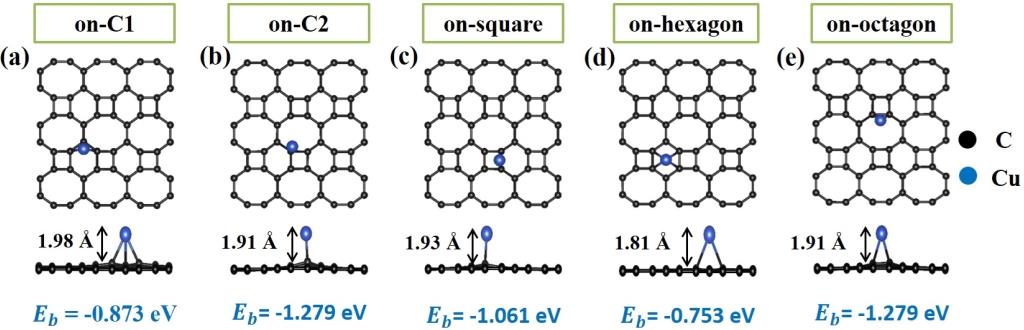}
	\caption{Optimized configurations of $\mathrm{CO_2}$ adsorbed over Cu-adsorbed pristine Biphenylene (Cu-BPH) SAC.} 
	\label{Fig. S6}
\end{figure*}

\begin{figure*}[b]
	\centering
	\includegraphics[width=\columnwidth,keepaspectratio]{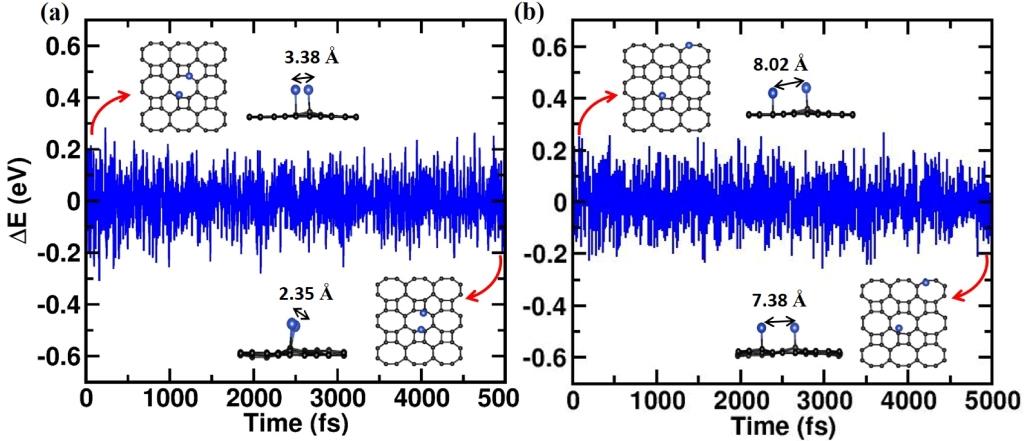}
	\caption{Molecular dynamics simulations indicating the evolution of energy change with time at 300 K, when the two Cu-atoms are placed (a) at a near distance of 3.38\AA\ and (b) at a far distance of 8.03 \AA\ on pristine BPH. Insets indicate the top and side views of the initial and final configurations. The distance between the two Cu-atoms is indicated by a arrow. The Cu and C-atom are indicated by blue and black spheres, respectively.} 
	\label{Fig. S7}
\end{figure*}

\begin{figure*}[hbt]
	\centering
	\includegraphics[width=\columnwidth,keepaspectratio]{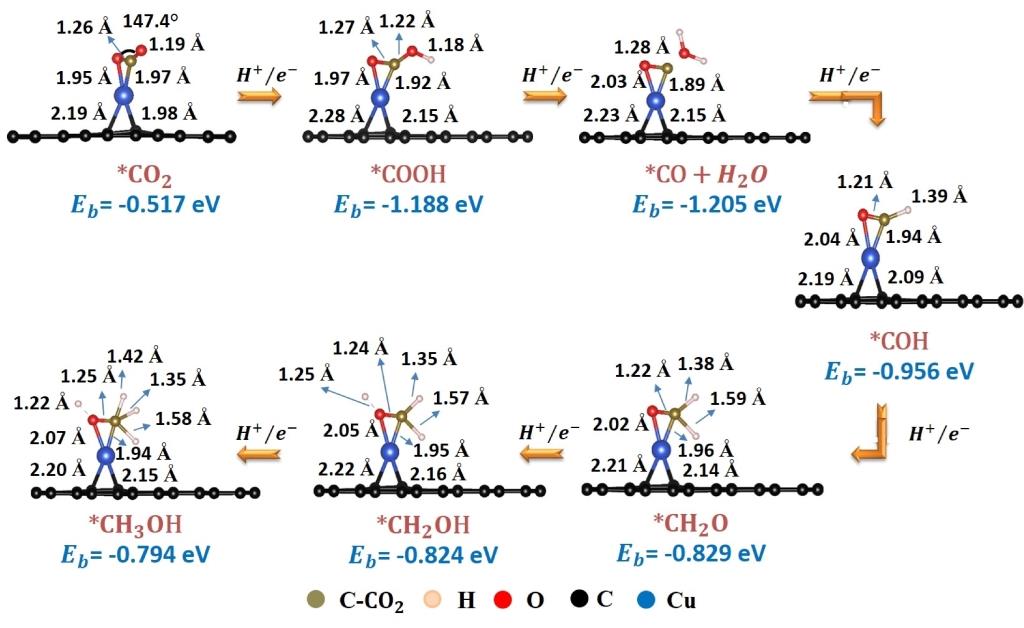}
	\caption{The electrocatalytic $\mathrm{CO_2}$ reduction reaction mechanism on Cu-BPH SAC under the effect of charged hydrogen (proton).} 
	\label{Fig. S8}
\end{figure*}

\FloatBarrier

	\bibliography{references.bib}